\documentclass[preprint,pre,groupedaddress,amsmath,amssymb,endfloats*]{revtex4}

\usepackage{mathrsfs}
\usepackage{graphicx}
\usepackage{epstopdf}
\usepackage{color}

\begin{document}

\title{Three-body critical Casimir forces}

\author{T. G. Mattos}
\email{tgmattos@des.cefetmg.br}
\affiliation{Departamento de F\'isica e Matem\'atica, Centro Federal de Educa\c c\~ao Tecnol\'ogica de Minas Gerais, Av. Amazonas 7675, 30510-000
Belo Horizonte, Brazil,}
\affiliation{Max-Planck-Institut f\"ur Intelligente Systeme, Heisenbergstr. 3,
D-70569 Stuttgart, Germany,}
\affiliation{IV. Institut f\"ur Theoretische Physik, Universit\"at
Stuttgart, Pfaffenwaldring 57, D-70569 Stuttgart, Germany}

\author{L. Harnau}
\affiliation{Max-Planck-Institut f\"ur Intelligente Systeme, Heisenbergstr. 3,
D-70569 Stuttgart, Germany,}
\affiliation{IV. Institut f\"ur Theoretische Physik, Universit\"at
Stuttgart, Pfaffenwaldring 57, D-70569 Stuttgart, Germany}

\author{S. Dietrich}
\affiliation{Max-Planck-Institut f\"ur Intelligente Systeme, Heisenbergstr. 3,
D-70569 Stuttgart, Germany,}
\affiliation{IV. Institut f\"ur Theoretische Physik, Universit\"at
Stuttgart, Pfaffenwaldring 57, D-70569 Stuttgart, Germany}

\newpage

\begin{abstract}
Within mean-field theory we calculate universal scaling functions associated with critical Casimir forces for a system consisting of three parallel cylindrical colloids immersed in a near-critical binary liquid mixture. For several geometrical arrangements and boundary conditions at the surfaces of the colloids we study the force between two colloidal particles in the direction normal to their axes, analyzing the influence of the presence of a third particle on that force. Upon changing temperature or the relative positions of the particles we observe interesting features such as a change of sign of this force caused by the presence of the third particle. We determine the three-body component of the forces acting on one of the colloids by subtracting the pairwise forces from the total force. The three-body contribution to the total critical Casimir force turns out to be more pronounced for small surface-to-surface distances between the colloids as well as for temperatures close to criticality. Moreover we compare our results with similar ones for other physical systems such as three atoms interacting via van der Waals forces.

% Keywords: Casimir effect, colloids, critical points, liquid mixtures, three-body effects
% 
% PACS: 82.70.Dd, 68.35.Rh, 82.60.-s, 05.40.-a
% 
% Colloids, Phase transitions and critical phenomena, Critical phenomena in physical chemistry, Fluctuation phenomena in statistical physics

\end{abstract}

\maketitle

\newpage

%-------------------------------------------------------------------------------

\section{Introduction}

%-------------------------------------------------------------------------------

The study of many-body systems typically starts from describing their interactions in terms of the sum of pair potentials. Often it is advantageous and transparent to analyze such systems via effective interactions which emerge via integrating out certain degrees of freedom. This procedure, however, often generates many-body forces, beyond the linear superposition of basic pairwise forces. This kind of nonadditivity is present in various systems such as colloidal suspensions~\cite{PhysRevLett.92.078301,PhysRevE.69.031402,nova6,B926845F,PhysRevE.87.020301,PhysRevE.88.042314,jcp:1.4819896,SoftMatter:10.9675}, systems governed by quantum-electrodynamic Casimir forces~\cite{PhysRevLett.99.080401,PhysRevA.77.030101,PhysRevA.80.022519,PhysRevLett.104.160402,
PhysRevA.83.042516,PhysRevLett.105.040403}, polymers~\cite{nova5,refId0,PhysRevE.64.021801,PhysRevE.83.061801,arXiv:1401.2064}, granular systems~\cite{RevModPhys.68.1259,PhysRevLett.96.178001,PhysRevLett.108.198001}, nematic colloids~\cite{miko}, and noble gases or nanoparticles with van der Waals forces acting among them~\cite{axilrod:299,Wells1983429,PhysRevLett.57.230,jakse:8504,jcp1.3432765,559484,jlt_cole,jcp1.4792137}.

Here we study quantitatively the importance of such many-body effects for critical Casimir forces (CCFs)~\cite{krech:1,brankov:1,1742-6596-161-1-012037}. These long-ranged effective forces arise upon approaching the critical point of the solvent due to the confinement of corresponding order parameter fluctuations~\cite{fisher_degennes} and have been analyzed by studying both theoretically and experimentally, the effective interaction between a single spherical colloid and a planar wall~\cite{PhysRevLett.74.3189,PhysRevB.51.13717,PhysRevLett.81.1885,Schlesener:JStatPhys.2003,eisenriegler:3299,kondrat:204902,nature_letters,epnews,PhysRevE.80.061143,PhysRevLett.101.208301,EPL:troendle.2009,trondle:074702,MolPhys:troendle.2011,C0SM00635A,PhysRevE.87.022130}, between two isolated spherical colloids~\cite{PhysRevLett.74.3189,PhysRevB.51.13717,PhysRevLett.81.1885,Schlesener:JStatPhys.2003,eisenriegler:3299}, as well as between two spherical colloids facing a homogeneous planar wall~\cite{galodoido}.

Recently Dang and co-authors~\cite{jcp:1.4819896} have used effective pair potentials between colloids, as inferred from experiments, in order to perform Monte Carlo simulations of the phase behavior of colloidal suspensions with near-critical solvents. They have observed that many-body effects become rather significant as the solvent is brought thermodynamically close to a critical point. They have also found that many-body effects tend to decrease the net attraction between colloids, which contrasts with previous results from mean-field calculations for a system of two spherical colloids facing a planar homogeneous substrate~\cite{galodoido}. In the latter case it was shown that many-body effects due to the presence of a substrate do not exhibit a uniform trend but can either increase or decrease the net attraction between colloids. In order to deepen the understanding of many-body effects concerning critical Casimir interactions we consider a system of three colloidal particles under the influence of CCFs. 

These forces are characterized by universal scaling functions, which depend on the geometry of the configuration, the thermodynamic state of the system, and the boundary condition for the order parameter at the surface of the colloids. In order to determine such universal scaling functions and in face of the complexity of the geometry we resort to mean-field theory (MFT), which captures these functions as the leading contribution to their systematic expansion in terms of $\epsilon =4-d$ where $d$ is the spatial dimension. To that end we consider the standard Landau-Ginzburg-Wilson Hamiltonian for critical phenomena, which is given by

\begin{equation}\label{LGW-Hamiltonian}
\mathscr{H}[\phi] = \int_V{\rm d}^d\mathbf{r} \left\{ \frac{1}{2}\left(
\nabla\phi \right)^2 + \frac{\tau}{2}\phi^2 + \frac{u}{4!}\phi^4 \right\} ~,
\end{equation}
\vspace{0.0000001cm}

\noindent
with appropriate boundary conditions (BCs). For the particular case of a binary liquid mixture near its demixing point, the order parameter $\phi(\mathbf{r},t)$ is proportional to the difference between the local concentration of one of the two components and its critical value. $V$ is the volume completely filled by the fluid, $\tau$ is proportional to the reduced temperature $t=(T-T_c)/T_c$, and the coupling constant $u>0$ stabilizes $\mathscr{H}$ for $t<0$. The bulk correlation length diverges upon approaching the bulk critical point as $\xi_{\pm}(t\rightarrow 0^{\pm}) = \xi_0^{\pm}\left| t\right|^{-\nu}$, where $\nu\simeq 0.63$ in $d=3$ and $\nu = 1/2$ in $d=4$, i.e., within MFT~\cite{Pelissetto2002549}. The two non-universal amplitudes $\xi_0^{\pm}$ are of molecular size and they form the universal ratio $\xi_0^+/\xi_0^-\approx 1.9$ for $d=3$ and $\xi_0^+/\xi_0^- = \sqrt{2}$ for $d=4$~\cite{phasetransition_privman}. The BCs account for the adsorption preference of the confining surfaces (in the present 
case, the surfaces of the three colloids) for one of the two species of the mixture. From an experimental point of view it is difficult to quantify the strength of the adsorption preference. However, for the case of colloidal polystyrene particles immersed in binary liquid mixtures of water and lutidine there is a rather pronounced adsorption preference~\cite{nature_letters,epnews,PhysRevE.80.061143,PhysRevLett.101.208301,EPL:troendle.2009,trondle:074702,MolPhys:troendle.2011}. Therefore we consider the critical adsorption fixed point~\cite{phasetransition_diehl} with the BC $\phi\arrowvert_{\rm surface} =\pm\infty$ at a particle surface, which we refer to as $(\pm)$. Renormalization group theory tells that for
$t\to 0$ the surface critical behavior of such systems is the one corresponding to this fixed point, irrespective of the actual strength of the adsorption preference for specific systems~\cite{phasetransition_diehl}.

The paper is organized as follows. In Sec.~\ref{section_the_system} we define the system and the scaling functions for the CCFs, as well as the normalization scheme. In Sec.~\ref{section_results} we present the numerical results obtained for the universal scaling functions of the CCFs, from which we extract and analyze the three-body effects. In Sec.~\ref{section_conclusion} we summarize our results and draw some conclusions.

%-------------------------------------------------------------------------------

\section{Physical system}\label{section_the_system}

%-------------------------------------------------------------------------------

We study the CCFs acting on three colloidal particles immersed in a near-critical binary liquid mixture, which is at the critical concentration. The surfaces of the colloids are considered to have a strong adsorption preference for one of the two components of the mixture corresponding to $(+)$ or $(-)$ BCs, respectively. We calculate the forces numerically within the aforementioned MFT [see Eq.~\eqref{LGW-Hamiltonian}].

\begin{figure} % FIG.1
\begin{center}
\includegraphics[width=0.65\textwidth]{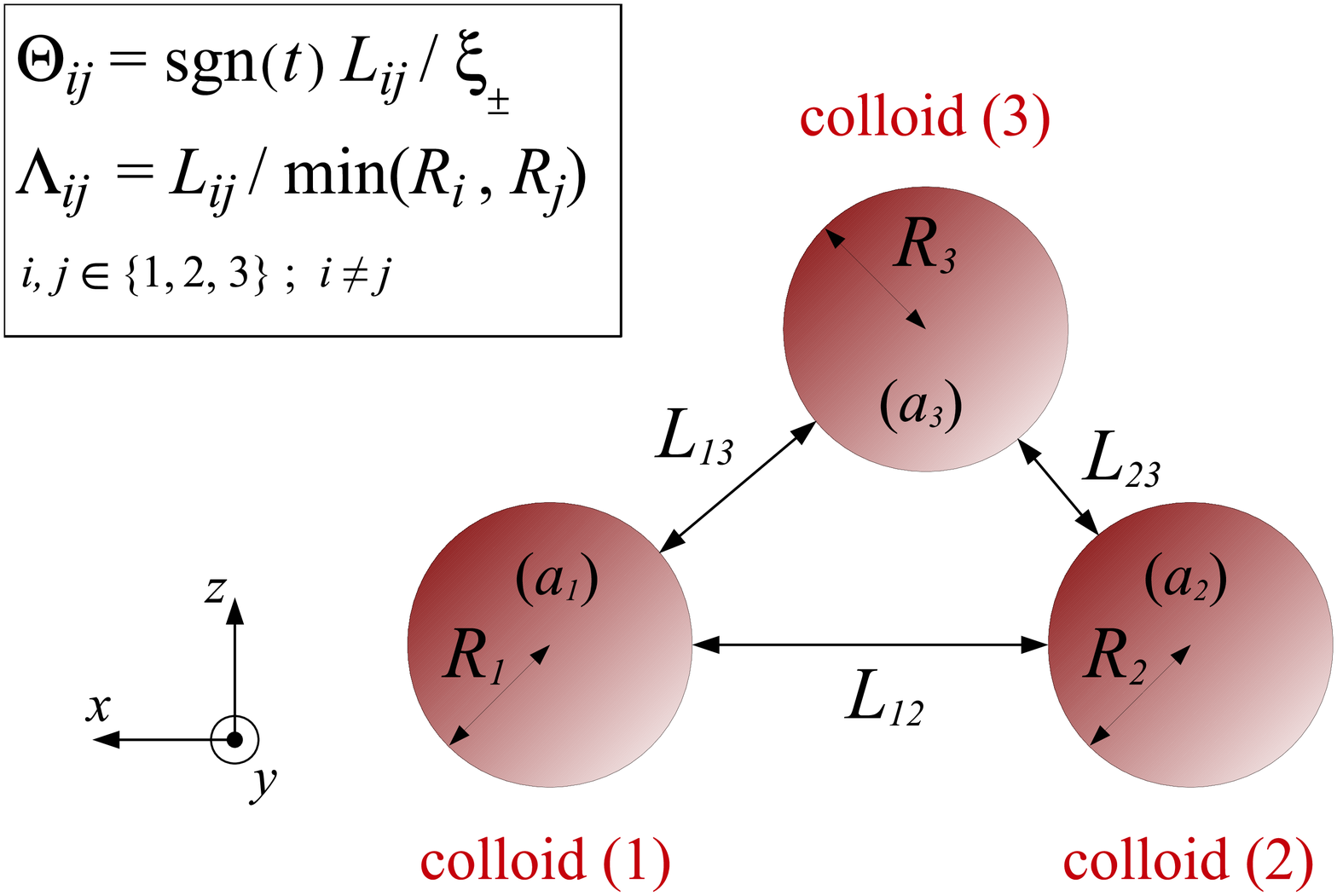}
\caption{
(Color online) Cross section of three parallel cylindrical colloidal particles of radii $R_i$ ($i=1,2,3$) with their axes parallel to the $y$-axis, immersed in a near-critical binary liquid mixture (not shown). The three colloidal particles with BCs $(a_i)$ are located at surface-to-surface distances $L_{ij}$, with $i,j\in \{1,2,3\}$ and $i\neq j$. In the case of four spatial dimensions the figure shows a cross section of the system, which is invariant also along the fourth direction, so that the cylinders correspond to parallel hypercylinders with two translationally invariant directions, i.e., the $y-$direction and the fourth dimension.}
\label{system_sketch}
\end{center} 
\end{figure}

In particular, we consider three three-dimensional, parallely aligned cylinders of radii $R_i$ with BCs $(a_i)$ at surface-to-surface distances $L_{ij}$, with $i,j\in \{1,2,3\}$ and $i\neq j$ (see Fig.~\ref{system_sketch}). In the context of CCFs such elongated particles have so far been investigated only in three theoretical studies~\cite{trondle:074702,PhysRevE.88.012137,C3SM52858H}. The BCs of the whole system are represented by the set $(a_1,a_2,a_3)$, where $a_1$, $a_2$, and $a_3$ can be either $+$ or $-$. It is important to mention that we discuss colloidal particles with the shape of hypercylinders in $d=4$ which are taken to be parallel along the $y$-direction as well as the fourth dimension with macroscopically long hyperaxes in these two directions. Considering such hypercylinders allows us to minimize $\mathscr{H}[\phi]$ numerically using a finite element method~\cite{f3dm} in order to obtain the spatially inhomogeneous order parameter profile $\phi(x,z)$ for the geometries under consideration. As compared with the case of spherical colloids, the analysis for cylindrical colloids is technically simpler because the system as a whole is translationally invariant along all directions but two, i.e., $x$ and $z$ (see Fig.~\ref{system_sketch}). This reduction in the number of relevant dimensions allows us to perform numerical calculations with adequate precision for a range of various model parameters which is wider than in the case of spherical colloids. It has advantageously turned out that qualitative and quantitative features of suitably normalized scaling functions of the CCF acting on a cylindrical colloid facing a homogeneous or inhomogeneous substrate are similar to the ones for a spherical colloid~\cite{trondle:074702}. For example the scaling functions for (++) BCs at the colloids and the substrate exhibit minima at $T > T_c$ which are, however, located closer to $T_c$ by a factor of 1.2 for spherical colloids as compared to cylindrical colloids of same radius. This quantitative difference is related to the fact that the decay rate of the scaling function of the order parameter profile for critical 
adsorption at colloids increases upon increasing the local curvature of the colloids~\cite{nova1}. Accordingly we anticipate that our present results also capture the corresponding trends for three spherical colloids.

%*******************************************

%*******************************************

We consider a binary liquid mixture at its critical concentration and with an upper critical demixing point. For this situation, $t > 0$ corresponds to the mixed, disordered phase of the fluid, while $t < 0$ corresponds to the demixed, ordered phase. The meaning of the sign has to be reversed if one considers a lower critical point.

We are interested in the CCF $F^{(1,x)}_{(a_1,a_2,a_3)}(L_{12},L_{13},L_{23},R_1,R_2,R_3,t)$ acting on colloid (1) in the presence of the colloids (2) and (3) along the $x$-direction. It takes the scaling form

\begin{equation} \label{def_scaling_1}
F^{(1,x)}_{(a_1,a_2,a_3)}(L_{12},L_{13},L_{23},R_1,R_2,R_3,t)=k_BT \frac{R_1}{L_{12}^d}
K^{(1,x)}_{(a_1,a_2,a_3)}(\Theta_{12},\Theta_{13},\Theta_{23},\Lambda_{12},\Lambda_{13},\Lambda_{23})\,,
\end{equation}
\vspace{0.0000001cm}

\noindent
where $\Lambda_{ij}=L_{ij}/{\rm min}(R_i,R_j)$ and $\Theta_{ij}={\rm sign}(t) L_{ij}/\xi_\pm$. Equation \eqref{def_scaling_1} describes the contribution to the CCF along the $x$-axis that emerges upon approaching $T_c$. $F^{(1,x)}$ is the force divided by the product of the lengths of the hypercylinder in its $d-2$ translationally invariant directions (see Fig.~\ref{system_sketch}). We analyze the scaling functions $K^{(1,x)}_{(a_1,a_2,a_3)}$ within MFT as given by Eq.~\eqref{LGW-Hamiltonian} for hypercylinders in $d=4$, which captures the correct scaling functions in $d=4$ up to logarithmic corrections occurring in the upper critical dimension $d_c$~\cite{phasetransition_diehl,0295-5075-74-1-022}, with $d_c = 4$ here.

As a reference system we consider \textit{two} parallel cylindrical colloids of the same size ($R_1=R_2=R$) with BCs $(a_1)$ and $(a_2)$ separated by a surface-to-surface distance $L$; their axes are parallel to the $y$-axis and their centers lie on the $x$-axis (see Fig.~\ref{system_sketch} in the absence of colloid (3)). Colloid (1) experiences a CCF, divided by the product of the lengths of the hypercylinders in their $d-2$ translationally invariant directions,

\begin{equation} \label{def_scaling_norm}
F^{(*,x)}_{(a_1,a_2)}(L,R,t) = k_B T \frac{R}{L^d}
K^{(*,x)}_{(a_1,a_2)}(\Theta={\rm sign}(t) \dfrac{L}{\xi_\pm},\Lambda=\dfrac{L}{R})~.
\end{equation}
\vspace{0.0000001cm}

\noindent
In the following we normalize the scaling functions $K^{(1,x)}_{(a_1,a_2,a_3)}$ by the quantity $K^{(*,x)}_{(+,+)}(\Theta=0,\Lambda=1)$, which corresponds to the amplitude of the CCF acting at $T_c$ on one of the two colloids with $(+,+)$ BCs at a surface-to-surface distance $L=R$. Therefore we consider the normalized scaling functions

\begin{equation}  \label{def_normalized_scaling}
\overline{K}^{(1,x)}_{(a_1,a_2,a_3)}(\Theta_{12},\Theta_{13},\Theta_{23},\Lambda_{12},\Lambda_{13},\Lambda_{23}) =
\frac{K^{(1,x)}_{(a_1,a_2,a_3)}(\Theta_{12},\Theta_{13},\Theta_{23},\Lambda_{12},\Lambda_{13},\Lambda_{23})}{K^{(*,x)}_{(+,+)}(\Theta=0,\Lambda=1)}.
\end{equation}
\vspace{0.0000001cm}

We note parenthetically that the standard normalization scheme uses the more easily accessible normalization amplitude $\Delta_{(+,+)}$ for the CCF at $T_c$ between two parallel plates with $(+,+)$ BCs, which is given within MFT by (see Ref.~\cite{trondle:074702} and references therein)

\begin{equation}
\Delta_{(+,+)} = -24\dfrac{[\mathcal{K}(1/\sqrt{2})]^4}{u} \simeq -283.61 / u ~,
\end{equation}
\vspace{0.0000001cm}

\noindent
where $\mathcal{K}$ is the elliptic integral of the first kind. Within MFT the amplitude $K^{(*,x)}_{(+,+)}(\Theta=0,\Lambda=1)$ can be expressed in terms of $\Delta_{(+,+)}$:

\begin{equation}\label{estaequacaoetaoimportantequeelaganhouseupropriolabel}
K^{(*,x)}_{(+,+)}(\Theta=0,\Lambda=1) \approx 0.2491 \times \Delta_{(+,+)} ~.
\end{equation}
\vspace{0.0000001cm}

\noindent
With Eq.~\eqref{estaequacaoetaoimportantequeelaganhouseupropriolabel} one is able to eliminate the coupling constant $u$, which remains unspecified within MFT.

\begin{figure} % FIG.2
\begin{center}
\includegraphics[width=0.65\textwidth]{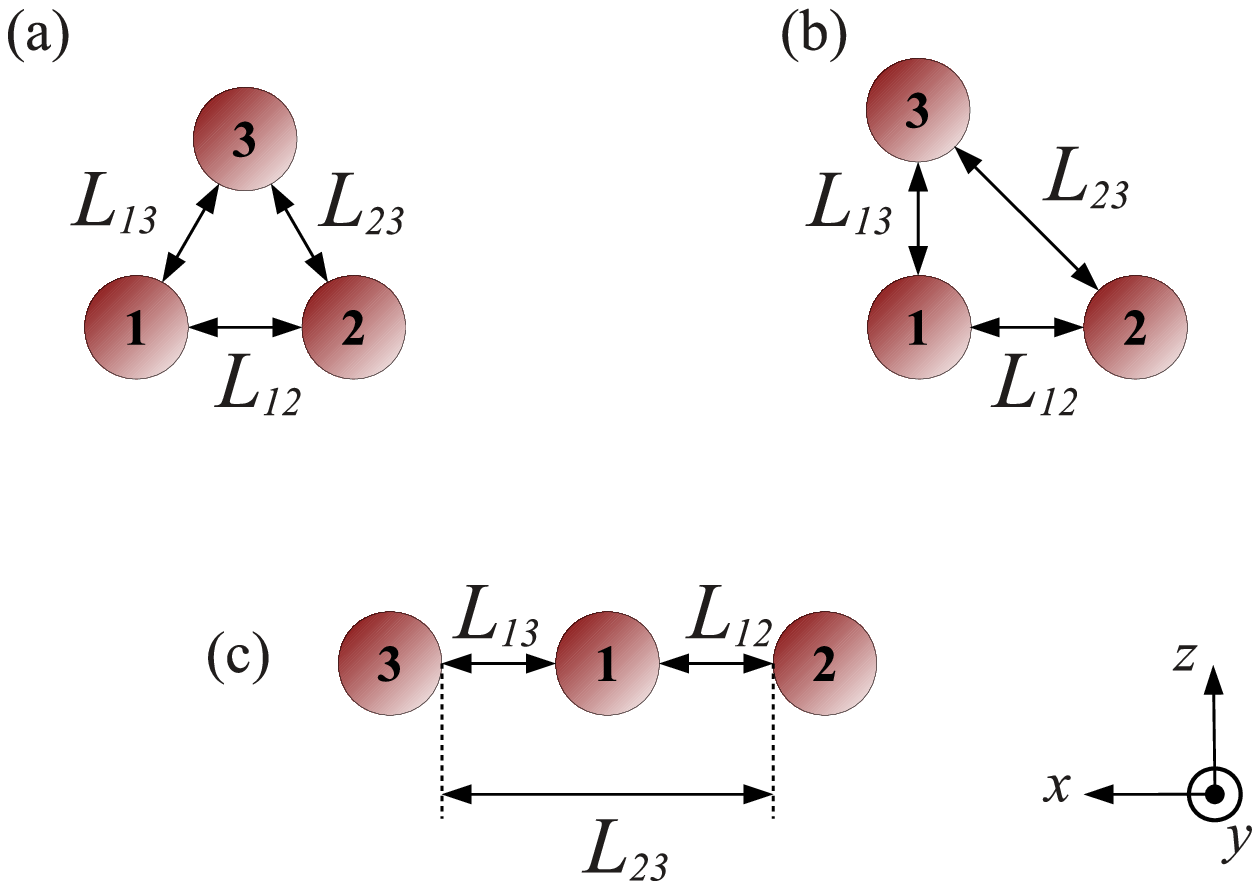}
\caption{
(Color online) Three configurations considered for three parallel cylindrical colloids: (a) isosceles triangle: $L_{13}=L_{23}$; (b) right-angled triangle: $(L_{23}+2R)^2=(L_{12}+2R)^2 + (L_{13}+2R)^2$; (c) line: $L_{23}=L_{12} + L_{13} + 2R$. All macroscopically extended cylindrical colloids are aligned parallel to the $y$-axis (see Fig.~\ref{system_sketch}).}
\label{arrangements}
\end{center} 
\end{figure}

We numerically determine the order parameter profiles $\phi(x,z)$ from which we calculate the CCFs by using the stress tensor~\cite{PhysRevE.56.1642,kondrat:204902,trondle:074702}. Specifically, we analyze the CCFs acting on colloid (1) for the configuration shown in Fig.~\ref{system_sketch}, for $t\geqslant 0$, and with the binary liquid mixture at its critical concentration. In the following we consider colloid (1) to have the fixed BC $(a_1=+)$ and equally sized colloids (i.e., $R_1=R_2=R_3=R$). We proceed by varying the surface-to-surface distances between the colloids by varying either $L_{12}, L_{13},$ or $L_{23}$. In particular we consider three special geometrical configurations for the three colloids (see Fig.~\ref{arrangements}): an isosceles triangle (i.e., $L_{13}=L_{23}$), a right-angled triangle with $(L_{23}+2R)^2=(L_{12}+2R)^2 + (L_{13}+2R)^2$, and a line with $L_{23}=L_{12} + L_{13} + 2R$. We also consider various sets of BCs for the colloids (2) and (3). In the following results the 
numerical 
error is typically less than $3\%$, unless explicitly stated otherwise.

%*******************************************

%*******************************************

%-------------------------------------------------------------------------------

\section{Results}\label{section_results}

%-------------------------------------------------------------------------------

\subsection{Isosceles triangle} \label{subsection_triangle}

\begin{figure} % FIG.3
\begin{center}
\includegraphics[width=0.65\textwidth]{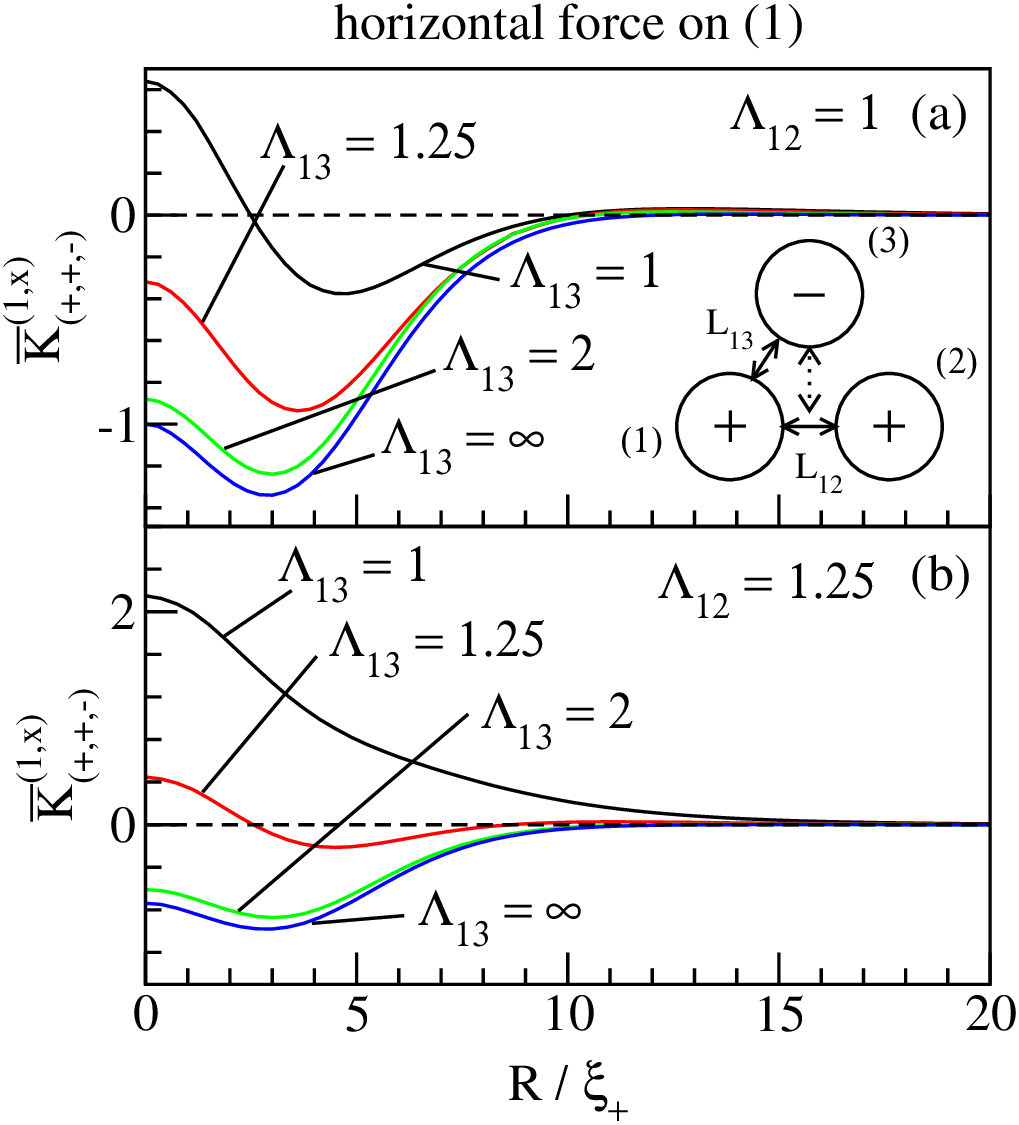}
\caption{
(Color online) Normalized scaling function
$\overline{K}^{(1,x)}_{(+,+,-)}(\Theta_{12}=L_{12}/\xi_+,\Lambda_{12}=L_{12}/R,\Lambda_{13}=L_{13}/R)$ of the CCF acting on colloid (1) along the $x$-direction (i.e., of the total horizontal force on (1)) for $\Lambda_{12}=1$ in (a) and $\Lambda_{12}=1.25$ in (b) with the centers of the colloids forming an isosceles triangle ($L_{13}=L_{23}$). The scaling function is shown for $t>0$ as function of the scaling variable ratio $R/\xi_+=\Theta_{12}/\Lambda_{12} = (|T - T_c| / T_c)^{\nu} R / \xi_0^+$ for four values of the scaling variable $\Lambda_{13}=L_{13}/R_3$: $\Lambda_{13}=1$ (black lines), $\Lambda_{13}=1.25$ (red lines), $\Lambda_{13}=2$ (green lines), and $\Lambda_{13}=\infty$ (blue lines), while $\Lambda_{23}=\Lambda_{13}$ for all curves in (a) and (b). For fixed $R$, each curve corresponds to a different vertical position of colloid (3) along the perpendicular bisector of $L_{12}$ (dotted double-headed arrow), with the colloids (1) and (2) fixed in space. As expected, upon increasing $\Lambda_{13}$ the scaling function approaches the blue lines, which correspond to the scaling function of the pairwise force between the colloids (1) and (2), i.e., if colloid (3) is infinitely far away from the colloids (1) and (2). $\overline{K}^{(1,x)}_{(+,+,-)}<0$ $(>0)$ implies that colloid (1) is attracted to (repelled from) colloid (2) along the $x$-direction. Note that according to our choice of the coordinate system the positive (negative) $x$-direction points away (towards) colloid (2) (see Figs.~\ref{system_sketch} and \ref{arrangements}). The circles and arrows in panel (a) are \textit{schematic} representations of the colloids and of the surface-to-surface distances between them; accordingly they are not drawn to scale.}
\label{triang__ppm_L12-fixed}
\end{center}
\end{figure}

\begin{figure} % FIG.4
\begin{center}
\includegraphics[width=0.65\textwidth]{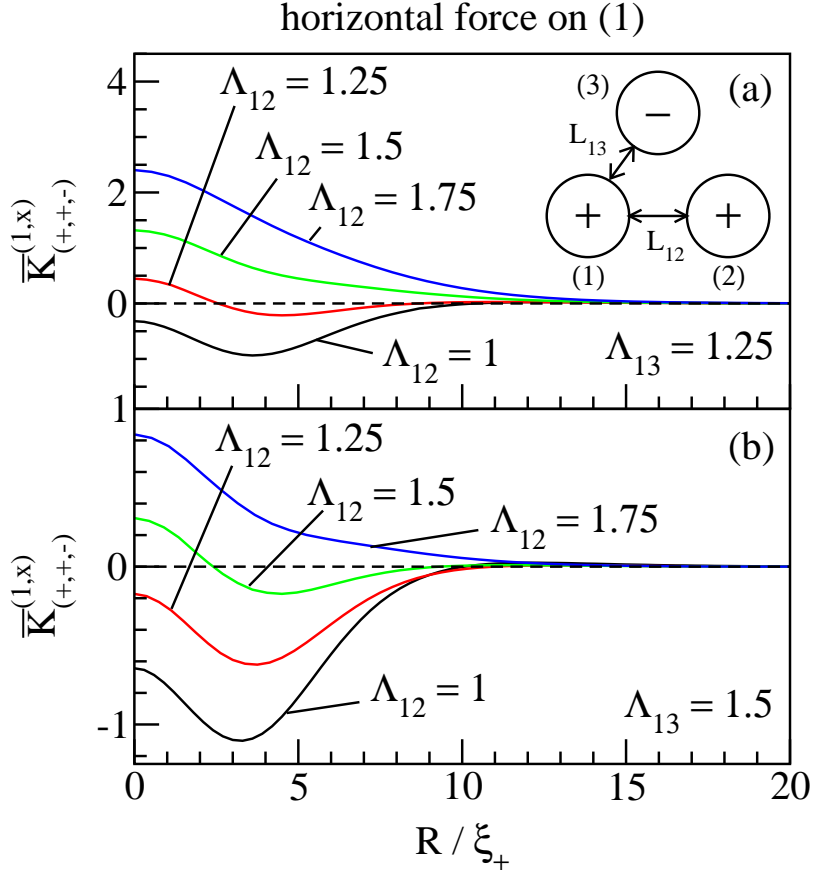}
\caption{
(Color online) Normalized scaling function
$\overline{K}^{(1,x)}_{(+,+,-)}(\Theta_{12}=L_{12}/\xi_+,\Lambda_{12}=L_{12}/R,\Lambda_{13}=L_{13}/R)$ of the total horizontal CCF acting on colloid (1) for $\Lambda_{13}=1.25$ in (a) and $\Lambda_{13}=1.5$ in (b) with the centers of the colloids forming an isosceles triangle ($L_{13}=L_{23}$). The scaling function is shown for $t>0$ as function of the scaling variable ratio $R/\xi_+ = \Theta_{12}/\Lambda_{12}$ for four values of the scaling variable $\Lambda_{12}=L_{12}/R$: $\Lambda_{12}=1$ (black lines), $\Lambda_{12}=1.25$ (red lines), $\Lambda_{12}=1.5$ (green lines), and $\Lambda_{12}=1.75$ (blue lines). For fixed $R$, the curves correspond to different surface-to-surface distances $L_{12}$ between colloids (1) and (2). Varying $L_{12}$ under the constraint $L_{13}=L_{23}$ implies that only colloid (3) is fixed in space and the centers of the colloids (1) and (2) move, equally but in opposite directions, on an arc centered at the center of colloid (3) and with radius $L_{13}+2R$. Accordingly, in this set up $\Lambda_{12}$ can vary between $0\leq\Lambda_{12}\leq 2\Lambda_{13}+2$ where the maximum value corresponds to a linear configuration with colloid (3) in the center. $\overline{K}^{(1,x)}_{(+,+,-)}<0$ $(>0)$ implies that colloid (1) is attracted to (repelled from) colloid (2) along the $x$-direction. The circles and arrows in panel (a) are \textit{schematic} representations of the colloids and of the surface-to-surface distances between them; accordingly they are not drawn to scale.} 
\label{triang__ppm_L13-fixed}
\end{center}
\end{figure}

\begin{figure} % FIG.5
\begin{center}
\includegraphics[width=0.65\textwidth]{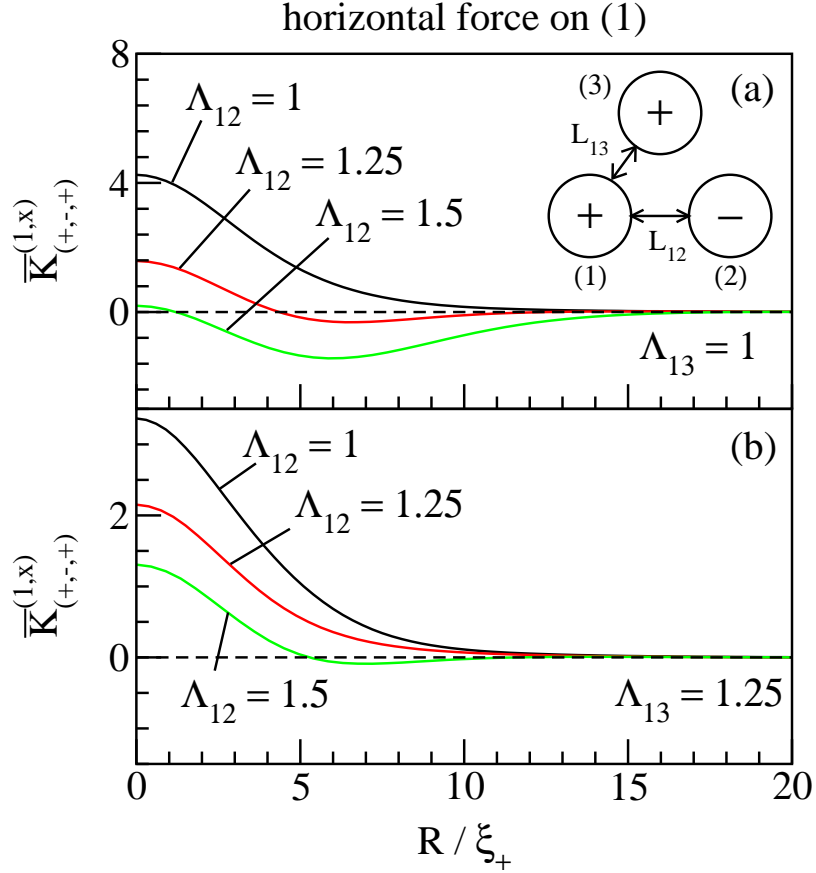}
\caption{
(Color online) Normalized scaling function
$\overline{K}^{(1,x)}_{(+,-,+)}(\Theta_{12}=L_{12}/\xi_+,\Lambda_{12}=L_{12}/R,\Lambda_{13}=L_{13}/R)$ of the total horizontal CCF acting on colloid (1) in the presence of the colloids (2) and (3). The scaling function is shown for $t>0$ as function of the scaling variable ratio $R/\xi_+=\Theta_{12}/\Lambda_{12}$ for $\Lambda_{13}=1$ in (a) and $\Lambda_{13}=1.25$ in (b) with the centers of the colloids forming an isosceles triangle ($L_{13} = L_{23}$). Each curve corresponds to a certain value of the scaling variable $\Lambda_{12}=L_{12}/R_1$: $\Lambda_{12}=1$ (black curves), $\Lambda_{12}=1.25$ (red curves), and $\Lambda_{12}=1.5$ (green curves). For fixed  $R$, the curves correspond to distinct surface-to-surface distances between colloids (1) and (2). Varying $L_{12}$ under the constraint $L_{13} = L_{23}$ implies that here one has the same kind of sequences of configurations as in Fig.~\ref{triang__ppm_L13-fixed}. $\overline{K}^{\,(1,x)}_{(+,-,+)}<0$ $(>0)$ implies that colloid (1) is attracted to (repelled from) colloid (2) in the $x$-direction. The circles and arrows in panel (a) are \textit{schematic} representations of the colloids and of the surface-to-surface distances between them; accordingly they are not drawn to scale.}
\label{triang__pmp_L13-fixed}
\end{center} 
\end{figure}

\begin{figure} % FIG.6
\begin{center}
\includegraphics[width=0.65\textwidth]{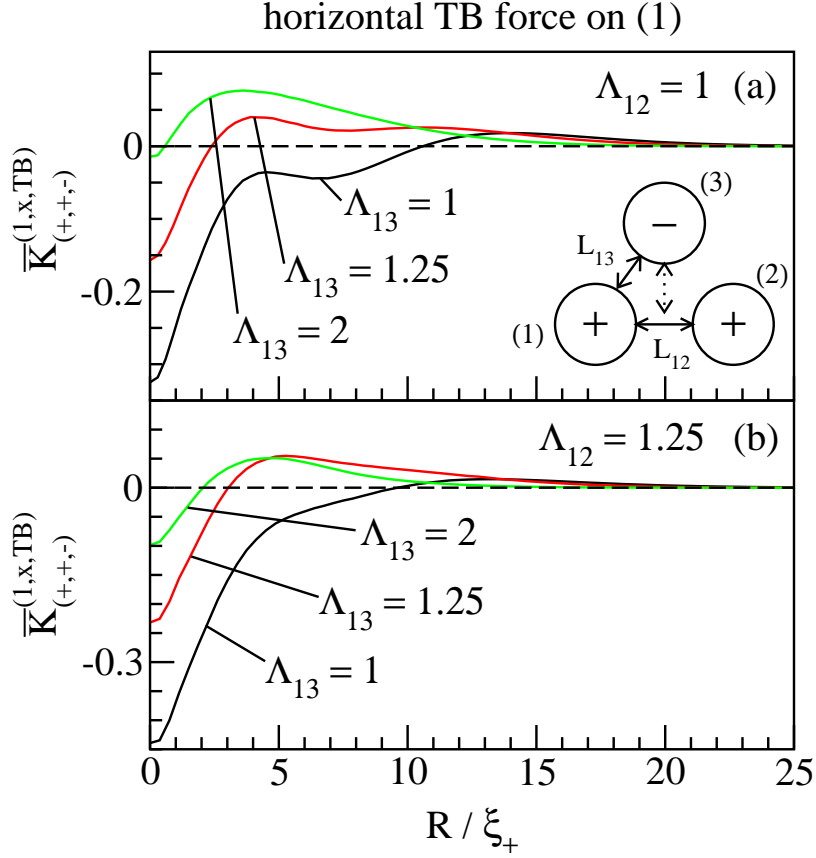}
\caption{
(Color online) Normalized scaling function
$\overline{K}^{(1,x,TB)}_{(+,+,-)}(\Theta_{12}=L_{12}/\xi_+,\Lambda_{12}=L_{12}/R,\Lambda_{13}=L_{13}/R)$ of the three-body horizontal CCF acting on colloid (1) for $\Lambda_{12}=1$ in (a) and $\Lambda_{12}=1.25$ in (b) with the centers of the colloids forming an isosceles triangle ($L_{13}=L_{23}$). The scaling function is shown as function of the scaling variable ratio $R/\xi_+ = \Theta_2/\Lambda_{12}$. The black, red, and green lines correspond to $\Lambda_{13}=1, \Lambda_{13}=1.25$, and $\Lambda_{13}=2$, respectively. For fixed $R$, each curve corresponds to a different vertical position of the colloid (3) along the perpendicular bisector of $L_{12}$ (dotted double-headed arrow), with the colloids (1) and (2) fixed in space. Figures \ref{triang__ppm_L12-fixed} and \ref{triang__ppm_TB_L12-fixed} provide a direct comparison between the total horizontal CCF and the corresponding three-body contribution (note the different scales of the coordinates). The circles and arrows in panel (a) are \textit{schematic} representations of the colloids and of the surface-to-surface distances between them; accordingly  are not drawn to scale.}
\label{triang__ppm_TB_L12-fixed}
\end{center} 
\end{figure}

First we consider a colloid configuration in which the centers of the colloids form the vertices of an isosceles triangle with $L_{13}=L_{23}$ [see Fig.~\ref{arrangements}~(a)]. In Fig.~\ref{triang__ppm_L12-fixed} we show the behavior of the normalized [Eq.~\eqref{def_normalized_scaling}] scaling function $\overline{K}^{(1,x)}_{(+,+,-)}(\Theta_{12}={\rm sign}(t)L_{12}/\xi_{\pm},\Lambda_{12}=L_{12}/R,\Lambda_{13}=L_{13}/R)$ of the  $x$-component of the CCF, i.e., the horizontal CCF acting on colloid (1) with $(a_1=+)$ BC and in the presence of the colloids (2) and (3) with $(a_2=+)$ and $(a_3=-)$ BCs, respectively. The scaling functions are shown as functions of the scaling variable ratio $\Theta_{12}/\Lambda_{12}=R/\xi_+$, i.e., for $t>0$. By using the relationship $R/\xi_+ = \left( | T-T_c |/T_c \right)^{\nu}R/\xi_0^+$ the scaling variable ratio $R/\xi_+$ used in Figs.~\ref{triang__ppm_L12-fixed}~$-$~\ref{ideia_fraca} can be expressed in terms of the temperature $T$. In Fig.~\ref{triang__ppm_L12-fixed}~(a) the surface-to-surface distance between the colloids (1) and (2) is $L_{12}=R$ (i.e., $\Lambda_{12}=1$), while in Fig.~\ref{triang__ppm_L12-fixed}~(b) $L_{12}=1.25R$ (i.e., $\Lambda_{12}=1.25$). The curves shown correspond to four values of $\Lambda_{13}=L_{13}/R$. For fixed $R$, Figs.~\ref{triang__ppm_L12-fixed}~(a) and (b) show the temperature dependence of the horizontal CCF acting on colloid (1) for four  values of $L_{13}=L_{23}$, with the colloids (1) and (2) fixed in space. The blue bottom lines correspond to the situation in which the third colloid is infinitely far away from the colloids (1) and (2), which is equivalent to the pairwise interaction between those colloids.

From Fig.~\ref{triang__ppm_L12-fixed} one can see that, upon increasing $L_{13}$ (or $\Lambda_{13}$) while keeping the distance $L_{12}$ between the colloids (1) and (2) fixed, the curves approach the blue ones, which correspond to the scaling function associated with the pairwise CCF between the colloids (1) and (2). One can also infer from Fig.~\ref{triang__ppm_L12-fixed} that, for fixed temperature and fixed $\Lambda_{12}$ (i.e., $\Theta_{12}/\Lambda_{12}$ fixed) there is a change of sign of the scaling function upon varying the position $\Lambda_{13}$ of colloid (3). This means that, for a given temperature and surface-to-surface distance between colloids (1) and (2) (i.e., fixed $\Theta_{12}$ and $\Lambda_{12}$), there is a position $\Lambda_{13}$ of colloid (3) above the midpoint of the line connecting the centers of the colloids (1) and (2), for which the horizontal CCF acting on colloid (1) is zero. Furthermore, for an equilateral triangle there is a change of sign of the scaling function upon varying the temperature, as one can infer from the black curve in Fig.~\ref{triang__ppm_L12-fixed}~(a) and from the red curve in Fig.~\ref{triang__ppm_L12-fixed}~(b), corresponding to $\Lambda_{12}=\Lambda_{13}= \Lambda_{23}=1$ and $\Lambda_{12}=\Lambda_{13}= \Lambda_{23}=1.25$, respectively.

%*******************************************

%*******************************************

Figure~\ref{triang__ppm_L13-fixed} shows the behavior of the normalized scaling function $\overline{K}^{(1,x)}_{(+,+,-)}(\Theta_{12}={\rm sign}(t)L_{12}/\xi_{\pm},\Lambda_{12}=L_{12}/R,\Lambda_{13}=L_{13}/R)$ of the horizontal CCF acting on colloid (1) in the presence of the colloids (2) and (3) for $\Lambda_{13}=\Lambda_{23}=1.25$ in (a) and $\Lambda_{13}=\Lambda_{23}=1.5$ in (b). The scaling functions are shown for $t>0$ as functions of the scaling variable ratio $R/\xi_+=\Theta_{12}/\Lambda_{12}$. The various lines correspond to certain values of the scaling variable $\Lambda_{12}=L_{12}/R$ which for fixed $R$ correspond to distinct values of the surface-to-surface distance $L_{12}$ between the colloids (1) and (2). From Fig.~\ref{triang__ppm_L13-fixed} one infers that, for a fixed temperature, there is a change of sign of the scaling function upon varying $L_{12}$ (or $\Lambda_{12}$), indicating the existence of an \textit{unstable} equilibrium configuration, because the force turns from attractive to repulsive upon increasing $\Lambda_{12}$. Hence, for colloid-colloid distances $\Lambda_{12}$ sufficiently large, colloid (1) is pushed away from colloid 2 due to the dominating repulsion between colloids (1) and (3) in spite of the attraction between the colloids (1) and (2), whereas for short distances $\Lambda_{12}$ it is pulled towards colloid (2) due to the attraction between them dominating the repulsion between the colloids (1) and (3).

%*******************************************

%*******************************************

In Fig.~\ref{triang__pmp_L13-fixed} we show the behavior of the normalized scaling function $\overline{K}^{(1,x)}_{(+,-,+)}(\Theta_{12}={\rm sign}(t)L_{12}/\xi_{\pm},\Lambda_{12}=L_{12}/R,\Lambda_{13}=L_{13}/R)$ of the horizontal CCF acting on colloid (1) for $\Lambda_{13}=\Lambda_{23}=1$ in (a) and $\Lambda_{13}=\Lambda_{23}=1.25$ in (b). The scaling function is shown as function of the scaling variable ratio $R/\xi_+=\Theta_{12}/\Lambda_{12}$ for three values of $\Lambda_{12}$: $\Lambda_{12}=1$ (black curves), $\Lambda_{12}=1.25$ (red curves), and $\Lambda_{12}=1.5$ (green curves). From Fig.~\ref{triang__pmp_L13-fixed}~(a) one infers that, for a fixed temperature, there is a \textit{stable} levitation position for colloid (1) along the $x$-direction: upon increasing the surface-to-surface distance between colloid (1) with $(a_1 = +)$ BC and colloid (2) with $(a_2 = -)$ BC, the scaling function turns from positive to negative, implying that the force changes from repulsive to attractive. In this case, the attraction between colloids (1) and (3) is dominating for large colloid-colloid distances $\Lambda_{12}$, while the repulsion between colloids (1) and (2) dominates for small values of $\Lambda_{12}$.
 
%*******************************************

%*******************************************

We have determined the bona fide three-body (TB) CCF $\mathbf{F}^{(1,TB)}_{(a_1,a_2,a_3)}$ acting on colloid (1) by subtracting from the total CCF $\mathbf{F}^{(1)}_{(a_1,a_2,a_3)}$ the sum of the pairwise forces acting on it:

\begin{equation}\label{resulting_TB_force}
\mathbf{F}^{(1,TB)}_{(a_1,a_2,a_3)} =
\mathbf{F}^{(1)}_{(a_1,a_2,a_3)} - \mathbf{F}^{(12)}_{(a_1,a_2)} - \mathbf{F}^{(13)}_{(a_1,a_3)}\, ,
\end{equation}
\vspace{0.0000001cm}

\noindent
where $\mathbf{F}^{(1j)}_{(a_1,a_j)}$ is the pairwise CCF acting on colloid (1) due to colloid ($j$) for $j=2,3$ and in the absence of the third colloid; $\mathbf{F}^{(12)}_{(a_1,a_2)} \cdot \mathbf{e}_x\equiv F^{(*,x)}_{(a_1,a_2)}$ (see Eq.~\eqref{def_scaling_norm}) where $\mathbf{e}_x$ is the unit vector pointing into the positive $x$-direction, i.e., away from colloid (2). In Eq.~\eqref{resulting_TB_force} the total force and the three-body force depend on all variables $L_{12}, L_{13}, L_{23}, R_1, R_2, R_3,$ and $t$, whereas $\mathbf{F}^{(1j)}$ depends only on $L_{1j}, R_1, R_j$, and $t$. We have studied the projection of $\mathbf{F}^{(1,TB)}_{(a_1,a_2,a_3)}$ onto the $x$-axis (see Fig.~\ref{system_sketch}):

\begin{equation}\label{def_TB_force}
F^{(1,x,TB)}_{(a_1,a_2,a_3)} = \mathbf{F}^{(1,TB)}_{(a_1,a_2,a_3)} \cdot \mathbf{e}_x\, .
\end{equation}
\vspace{0.0000001cm}

\noindent
The three-body CCF in Eq.~\eqref{def_TB_force} is characterized by the corresponding scaling function [compare Eq.~\eqref{def_scaling_1}]:

\begin{equation} \label{def_scaling_TB}
F^{(1,x,TB)}_{(a_1,a_2,a_3)}=k_BT \frac{R_1}{L_{12}^d}
K^{(1,x,TB)}_{(a_1,a_2,a_3)}(\Theta_{12},\Theta_{13},\Theta_{23},\Lambda_{12},\Lambda_{13},\Lambda_{23}).
\end{equation}
\vspace{0.0000001cm}

As before we consider here the special case  $R_1=R_2=R_3\equiv R$ and $L_{13} = L_{23}$. Moreover we introduce the relative contribution of the three-body component of the total CCF as

\begin{equation} \label{relative_contribution}
\delta = \frac{\left|F^{(1,x,TB)}\right|}{\left|F^{(1,x,TB)}\right|+\left|F^{(12,x)}\right|+\left|F^{(13,x)}\right|}\, ,
\end{equation}
\vspace{0.0000001cm}

\noindent
where $F^{(1j,x)} = \mathbf{F}^{(1j)}\cdot \mathbf{e}_x$.

In Fig.~\ref{triang__ppm_TB_L12-fixed} we show the normalized [see Eqs.~\eqref{def_scaling_norm} and \eqref{def_normalized_scaling}] scaling function $\overline{K}^{(1,TB)}_{(+,+,-)}(\Theta_{12}={\rm sign}(t)L_{12}/\xi_{\pm},\Lambda_{12}=L_{12}/R,\Lambda_{13}=L_{13}/R)$ of the three-body contribution to the total horizontal CCF acting on colloid (1). As one can infer from Fig.~\ref{triang__ppm_TB_L12-fixed}, the three-body contribution to the total CCF can be either positive (i.e., repulsive) or negative (i.e., attractive), depending both on temperature and geometry. Thus for a given value of $\Lambda_{13}$ (i.e., along one of the curves shown in Fig.~\ref{triang__ppm_TB_L12-fixed}) there is a temperature for which the three-body contribution to the total horizontal CCF acting on colloid (1) is zero. Hence under such a condition the sum of pairwise forces provides a quantitatively reliable description of the total horizontal force acting on colloid (1). For temperatures sufficiently far from $T_c$, the three-body CCF is always repulsive and decays exponentially for $R/\xi_+\to \infty$. The relative contribution of the three-body component of the total CCF reaches $10\%$ for $\Lambda_{12}=\Lambda_{13}=1$ and, as expected, this contribution decreases upon increasing $\Lambda_{13}$ with $\Lambda_{12}$ fixed.

%*******************************************

%*******************************************

We discuss our results by putting them into the context of similar ones obtained for other systems. First, we consider the van der Waals interaction between three atoms as described by the Axilrod-Teller three-atom potential~\cite{axilrod:299}

\begin{equation}\label{axilrod_teller_3bpot}
\mathcal{U}_{123} = C\frac{3\cos{\gamma_1}\cos{\gamma_2}\cos{\gamma_3} + 1}{r_{12}^3r_{23}^3r_{31}^3}\, ,
\end{equation}
\vspace{0.0000001cm}

\noindent
where $r_{ij}=\sqrt{\left( x_i - x_j \right)^2 + \left( y_i - y_j \right)^2 + \left( z_i - z_j \right)^2} = r_{ji}$ with $i,j\in \{ 1,2,3\}$ are the center-to-center distances between the atoms located at $(x_i, y_i, z_i)$; $\gamma_1, \gamma_2, \gamma_3$ are the angles formed by the lines along $r_{12}$ and $r_{13}$, $r_{21}$ and $r_{23}$, and $r_{32}$ and $r_{31}$, respectively, with $\gamma_1 + \gamma_2 + \gamma_3 = 180^{\circ}$. Without loss of generality, we choose $y_1=y_2=y_3=z_1=z_2=x_2=0$ and $x_1, x_3, z_3 >0$ for geometrical atom configurations corresponding to the ones for colloids as shown in Fig.~\ref{arrangements}. The coefficient $C$ is given by

\begin{equation}\label{equacaodocu}
C = \frac{3\hbar}{\pi}\int_0^{\infty} \alpha_1(i\omega) \alpha_2(i\omega) \alpha_3(i\omega) d\omega > 0 \, ,
\end{equation}
\vspace{0.0000001cm}

\noindent
where $\alpha_i > 0$ is the frequency-dependent polarizability of the atom of type $i=1,2,3$ (with the dimension of a volume). The force $\mathcal{F}^{(1,x)}_{123}$ due to the potential $\mathcal{U}_{123}$, which is the analogue of the three-body CCF, follows from differentiating $\mathcal{U}_{123}$ with respect to $x_1$~\cite{tesemestradoqquer}:

% \begin{equation}\label{equacaodocudentrodooutro}
% \mathcal{F}^{(1,x)}_{123} = -\frac{\partial}{\partial x_1}\mathcal{U}_{123} = -3\, \mathcal{U}_{123}\left( \frac{x_3 - x_1}{r_{31}^2} + \frac{x_2 - x_1}{r_{12}^2} \right)\, .
% \end{equation}
% \vspace{0.0000001cm}

\begin{equation}\label{equacaodocudentrodooutro}
\mathcal{F}^{(1,x)}_{123} = -\frac{\partial}{\partial x_1}\mathcal{U}_{123}\, .
\end{equation}
\vspace{0.0000001cm}

\noindent
If the atoms are arranged as an equilateral triangle [i.e., if $\gamma_1=\gamma_2=\gamma_3=60^{\circ}$ so that $\cos \gamma_i = 1/2$ or equivalently if $r_{12}=r_{23}=r_{31}$, which corresponds to $L_{12}=L_{23}=L_{31}$ with $R\to 0$ in Fig.~\ref{arrangements}~(a)], the three-atom force is

\begin{equation}\label{superjoia}
\mathcal{F}^{(1,x)}_{123} = \frac{99\,C}{16\,x_1^{10}}\, ,
\end{equation}

\noindent
which is positive, corresponding to repulsion, i.e., atom (1) is pushed away from atom (2) along the $x$-axis. In order to compare this with our results for three colloids one should consider the case of all BCs being equal, i.e., the $(+,+,+)$ BCs case. To this end we have determined the three-body component of the CCF acting on colloid (1) along the $x$-direction for the arrangement shown in Fig.~\ref{arrangements}~(a) for $\Lambda_{12}=\Lambda_{13}=\Lambda_{23}=$1, 1.25, and 1.5 with $(+,+,+)$ BCs. The result is qualitatively the same as in the case of the van der Waals interaction, i.e., colloid (1) is pushed away from colloid (2) along the $x$-axis. We emphasize that this van der Waals force has been derived for the case $x_1, x_3>0$ without loss of generality. For geometrical configurations with $x_1, x_3<0$, a similar calculation leads to a change of sign of the force in Eq.~\eqref{superjoia}, still corresponding to a repulsive three-atom force acting on atom (1).

We also compare our results with those obtained for the interaction between charged colloids. To this end we consider the results of direct measurements of the three-body interaction between charged colloids obtained by Brunner et al.~\cite{PhysRevLett.92.078301}. The system investigated by the authors consists of three equally sized charged colloids arranged as an isosceles triangle. The authors found that, regardless of the specific geometrical parameters of the arrangement, the three-body component of the force acting on the colloid corresponding to colloid (1) in Fig.~\ref{arrangements}~(a) along the line connecting colloids (1) and (2) is attractive [i.e., it pulls colloid (1) towards colloid (2)], while  the corresponding two-body force is repulsive, in agreement with earlier theoretical considerations using the nonlinear Poisson-Boltzmann theory~\cite{PhysRevE.66.011402}. Our results, on the other hand, show that the three-body component of the CCF can be either attractive or repulsive for $(+,+,+)$ BCs, depending on the 
geometrical parameters of the arrangement. In particular, for an arrangement corresponding to an equilateral triangle, the three-body component of the CCF acting on colloid (1) along the $x$-direction is always repulsive whereas the corresponding two-body CCF is attractive.

\subsection{Right-angled triangle} \label{subsection_L}

\begin{figure} % FIG.7
\begin{center}
\includegraphics[width=0.65\textwidth]{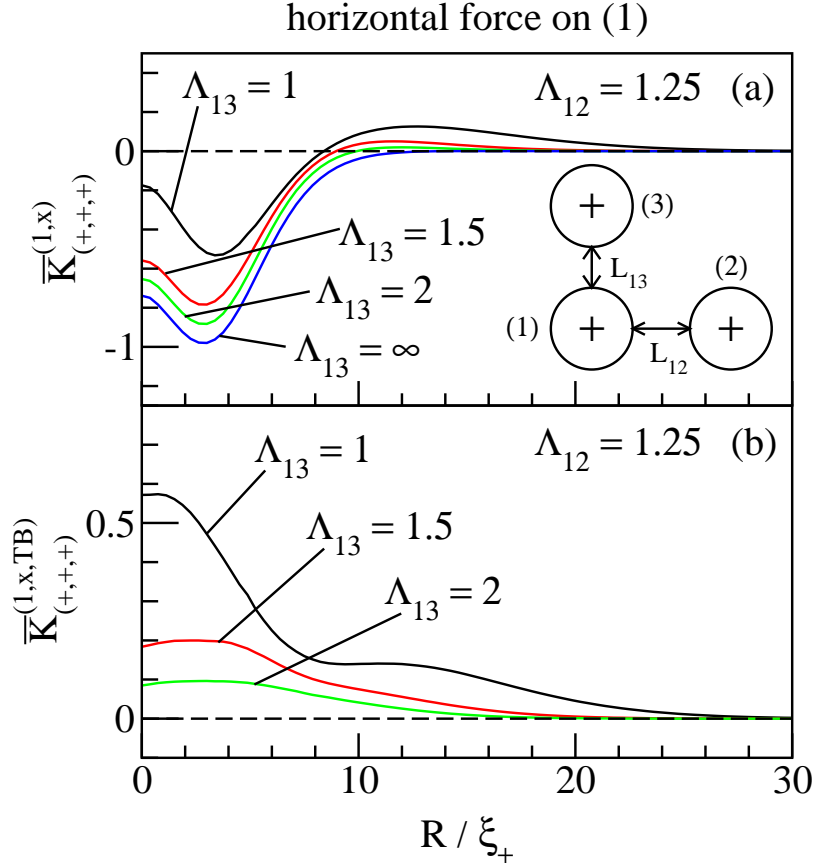}
\caption{
(Color online) Normalized scaling function $\overline{K}^{(1,x)}_{(+,+,+)}(\Theta_{12}=L_{12}/\xi_+,\Lambda_{12}=L_{12}/R,\Lambda_{13}=L_{13}/R)$ in (a) and  $\overline{K}^{(1,x,TB)}_{(+,+,+)}(\Theta_{12},\Lambda_{12},\Lambda_{13})$ in (b) of the horizontal total CCF and the three-body CCF, respectively, acting on colloid (1) for a right-angled triangle, i.e.,  $(L_{23}+2R)^2=(L_{12}+2R)^2 + (L_{13}+2R)^2$ and $\Lambda_{12}=1.25$. The scaling functions are shown as functions of the scaling variable ratio $R/\xi_+=\Theta_2/\Lambda_{12}$. The black, red, green, and blue lines correspond to $\Lambda_{13}=1, \Lambda_{13}=1.5, \Lambda_{13}=2$, and $\Lambda_{13}=\infty$, respectively. In (b) the horizontal dashed line corresponds to $\Lambda_{13}=\infty$. The circles and arrows in panel (a) are \textit{schematic} representations of the colloids and of the surface-to-surface distances between them; accordingly they are not drawn to scale.}
\label{L__ppp_L12-fixed}
\end{center} 
\end{figure}

We now consider a configuration consisting of a right-angled triangle, i.e., for $(L_{23}+2R)^2=(L_{12}+2R)^2 + (L_{13}+2R)^2$ [see Fig.~\ref{arrangements}(b)]. In Fig.~\ref{L__ppp_L12-fixed}~(a) we show the normalized scaling functions $\overline{K}^{(1,x)}_{(+,+,+)}(\Theta_{12}={\rm sign}(t)L_{12}/\xi_{\pm},\Lambda_{12}=L_{12}/R,\Lambda_{13}=L_{13}/R)$ of the total horizontal CCF acting on colloid (1), for $\Lambda_{12}=L_{12}/R_1=1.25$. The various line colors correspond to distinct values of the scaling variable $\Lambda_{13}=L_{13}/R_3$. As one can infer from Fig.~\ref{L__ppp_L12-fixed}~(a), upon increasing the surface-to-surface distance $L_{13}$, the scaling function uniformly approaches the one corresponding to the pairwise CCF between colloids (1) and (2) (i.e., the blue curve). Furthermore one can infer that, for fixed values of the temperature sufficiently away from $T_c$, there is a change of sign of $\overline{K}^{(1,x)}_{(+,+,+)}$ upon varying the surface-to-surface distance between colloids (1) and (3): upon increasing $L_{13}$ and $R/\xi_+$ large enough the force changes from repulsive to attractive. A change of sign also takes place for a change of temperature for certain spatially fixed configurations [see the black curve in Fig.~\ref{L__ppp_L12-fixed}~(a)].

In Fig.~\ref{L__ppp_L12-fixed}~(b) we show the normalized scaling functions $\overline{K}^{(1,x,TB)}_{(+,+,+)}(\Theta_{12}={\rm sign}(t)L_{12}/\xi_{\pm},\Lambda_{12}=L_{12}/R,\Lambda_{13}=L_{13}/R)$ of the three-body component of the horizontal CCF acting on colloid (1) for the same configurations as in Fig.~\ref{L__ppp_L12-fixed}~(a), i.e., for $\Lambda_{12}=1.25$ and $\Lambda_{13}=1$ (black line), 
$\Lambda_{13}=1.5$ (red line), and $\Lambda_{13}=2$ (green line). As expected, the contribution of the three-body component of the CCF decreases upon increasing the surface-to-surface distance between colloids (1) and (3). Moreover, we note that this result for the three-body component of the horizontal CCF acting on colloid (1) contrasts with previous results obtained for three-body effects for CCF near a wall. In the case of two colloidal particles facing a homogeneous planar substrate at equal distances it was shown that in the case of all BCs being equal, i.e., if both colloids and the substrate exhibit $(+)$ BCs, the three-body component of the lateral CCF acting on one of the colloids is attractive (i.e., it pushes one colloid towards the other) if the surface-to-surface distances between the colloids and between the colloids and the substrate are equal to the radius common to the colloids (see Figs. 1 and 8 in Ref.~\cite{galodoido}). On the other hand, in the case of three colloids arranged as a right-angled triangle we have found that the three-body component of the horizontal CCF is positive and therefore repulsive, i.e., it pushes colloid (1) away from colloid (2). Since the substrate can be viewed as the surface of a very large colloid, this observation tells that the nature of the many-body CCF depends also sensitively on the relative sizes of the colloids.

%*******************************************

%*******************************************

We compare our results with those for a system of three atoms arranged as a right-angled triangle (see Fig.~\ref{arrangements}~(b) with $R\to 0$) interacting via a van der Waals potential. One obtains a right-angled triangle by setting $\gamma = \pi/2$, $x_3 = x_1$, and $x_2 > x_1$ so that the three-atom force $\mathcal{F}^{(1,x)}_{123}$ is

\begin{equation}
\mathcal{F}^{(1,x)}_{123} = \frac{3C}{x_1^4 \left( x_1^2 + z_3^2\right)^{5/2} z_3}\, ,
\end{equation}

\noindent
which is positive, meaning that the interaction is repulsive, i.e., atom (1) is pushed away from atom (2). This is also the case for a right-angled triangle arrangement of colloids with all BCs being equal, i.e., $(+,+,+)$, as can be seen in Fig.~\ref{L__ppp_L12-fixed}~(b).

We also compare our results with those obtained for driven granular mixtures~\cite{RevModPhys.68.1259}, which exhibit Casimir-like, long-ranged effective interactions~\cite{PhysRevLett.96.178001}. These systems consist of a set of ``intruder'' particles immersed in a uniformly agitated granular fluid (i.e., a set of granular particles which are typically ten times smaller than the intruders and which undergo inelastic binary collisions). Recently Shaebani et al. have calculated the forces acting on three intruders in such a granular mixture~\cite{PhysRevLett.108.198001}. Considering a right-angled triangle configuration they have found that the three-body component of the total force, acting on the particle corresponding to colloid (1) in Fig.~\ref{arrangements}~(b), along the line which connects the centers of those intruders which are the analogues of the colloids (1) and (2) in Fig.~\ref{arrangements}~(b) is attractive, while the corresponding two-body force between the colloids (1) and (2) is repulsive. The total force acting on particle (1) is smaller than the vectorial sum of the pairwise forces due to the presence of particles corresponding to colloids (2) and (3). These results differ qualitatively from the ones for the CCFs shown in Fig.~\ref{L__ppp_L12-fixed}~(b), where one can see that the three-body component of the horizontal CCF acting on colloid (1) is repulsive whereas the corresponding two-body force [blue line in Fig.~\ref{L__ppp_L12-fixed} (a)] is attractive. Hence for both systems the sign of the three-body force acting on particle (1) is opposite to the one of the two-body force. However, the relative contribution $\delta$ of the three-body component of the force [see Eq.~\eqref{relative_contribution}] is significantly larger in the case of the CCFs as compared to the granular system considered in Ref.~\cite{PhysRevLett.108.198001}: in the present case, it is about $24\%$ for $\Lambda_{12}=1.25$ and $\Lambda_{13}=1$ (black curve in Fig.~\ref{L__ppp_L12-fixed}), and $10\%$ for $\Lambda_{12}=1.25$ and $\Lambda_{13}=1.5$ (red curve in Fig.~\ref{L__ppp_L12-fixed}); in the case of the granular system, for a configuration of 
the intruders corresponding to $\Lambda_{12}=\Lambda_{13}=1$, the relative contribution of the three-body component is around $3\%$.

\subsection{Line} \label{subsection_line}

\begin{figure} % FIG.8
\begin{center}
\includegraphics[width=0.65\textwidth]{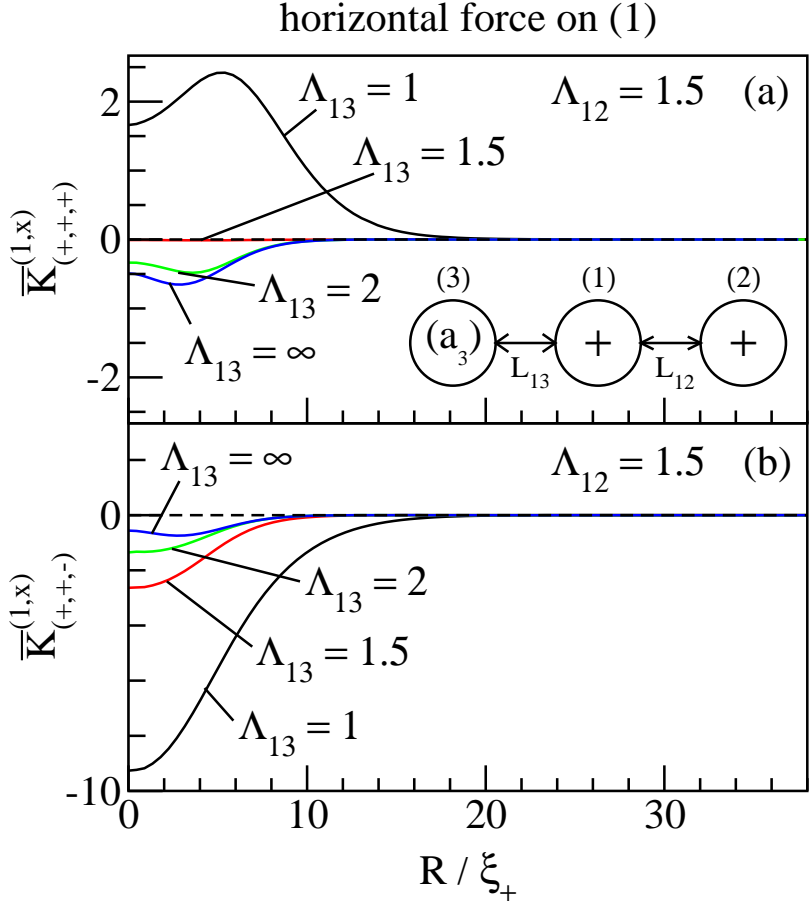}
\caption{
(Color online) Normalized scaling function $\overline{K}^{(1,x)}_{(+,+,a_3)}(\Theta_{12}=L_{12}/\xi_+,\Lambda_{12}=L_{12}/R,\Lambda_{13}=L_{13}/R)$ of the total CCF acting on colloid (1) for two linear configurations with $\Lambda_{12}=1.5$ and for $(a_3 = +)$ in (a) and $(a_3 = -)$ in (b). The centers of the three colloids lie on the $x$-axis so that $L_{23}=L_{12}+L_{13}+2R$ [see Fig.~\ref{arrangements}~(c)]. The scaling functions are shown as functions of the scaling variable ratio $R/\xi_+=\Theta_2/\Lambda_{12}$. The black, red, green, and blue lines correspond to $\Lambda_{13}=1, \Lambda_{13}=1.5, \Lambda_{13}=2$, and $\Lambda_{13}=\infty$, respectively. The circles and arrows in panel (a) are \textit{schematic} representations of the colloids and of the surface-to-surface distances between them; accordingly they are not drawn to scale.}
\label{line__ppp_L12-fixed}
\end{center} 
\end{figure}

Finally, we consider the situation in which the three colloids are aligned linearly, with their centers lying on the $x$-axis so that $L_{23}=L_{12}+L_{13}+2R$ [see Fig.~\ref{arrangements}~(c)]. In Fig.~\ref{line__ppp_L12-fixed} we show the normalized scaling functions $\overline{K}^{(1,x)}_{(+,+,a_3)}(\Theta_{12}=L_{12}/\xi_+,\Lambda_{12}=L_{12}/R,\Lambda_{13}=L_{13}/R)$ of the total CCF acting on colloid (1) along the $x$-direction, for $\Lambda_{12}=L_{12}/R_1=1.25$ and for $(a_3 = +)$ in (a) and $(a_3 = -)$ in (b). (Note that colloid (3) is the outer left one.) Each curve corresponds to a certain value of $\Lambda_{13}=L_{13}/R_3$: $\Lambda_{13}=1$ (black lines), $\Lambda_{13}=1.5$ (red lines), $\Lambda_{13}=2$ (green lines), and $\Lambda_{13}=\infty$ (blue lines). As one can see from Fig.~\ref{line__ppp_L12-fixed}, upon increasing $\Lambda_{13}$ [i.e., upon moving the left colloid (3) away from the colloids (1) and (2)] the shapes of the scaling functions $\overline{K}^{(1,x)}_{(+,+,a_3)}$ approach the same one which corresponds to the pairwise CCF for $(+,+)$ (blue lines, $\Lambda_{13}=\infty$). The red curve in Fig.~\ref{line__ppp_L12-fixed}(a) corresponds to the special case $L_{12}=L_{13}$ for which, due to all BCs being equal, the CCF acting on (1) is zero. As can be inferred from Fig.~\ref{line__ppp_L12-fixed}~(a), for every fixed value of the temperature, there is a change of sign in the total CCF upon changing the position of colloid (3), implying the existence of a \textit{stable} equilibrium position for colloid (1), which corresponds to $L_{12}=L_{13}$ [i.e., with the center of colloid (1) exactly at the midpoint of the line connecting the centers of colloids (2) and (3)]. From Fig.~\ref{line__ppp_L12-fixed}~(b) one can see that even for a surface-to-surface distance between the colloids (1) and (3) which is two times larger than the radius of the particles (i.e., for $\Lambda_{13}=2$), close to $T_c$ there remains a significant deviation from the pairwise force between the colloids (1) and (2). We attribute this behavior to the fact that the ratio of the strengths of the two-body CCFs for $(+,-)$ and $(+,+)$ BCs varies as a function of temperature~\cite{C3SM52858H}. Whereas close to $T_c$ the CCF for $(+,-)$ BCs is much stronger than the attractive CCF for $(+,+)$ BCs, both become comparable in strength for $\Theta_{12}\gg 1$. Therefore, even for $\Lambda_{13}=2$ the repulsive interaction between the colloids (1) and (3) significantly contributes to the force $\overline{K}^{(1,x)}_{(+,+,-)}$ on the colloid (1) shown in Fig.~\ref{line__ppp_L12-fixed}~(b) close to $T_c$.

%*******************************************

%*******************************************

In order to compare our results with those for analogous systems, we consider again the Axilrod-Teller three-atom potential given by Eq.~\eqref{axilrod_teller_3bpot}. For the case of three atoms in a line one can infer that the corresponding three-atom force is

\begin{equation}
\mathcal{F}^{(1,x)}_{123} = 6C\frac{2x_1 - x_3}{x_1^4\left( x_1-x_3 \right)^4 x_3^3}\, ,
\end{equation}

\noindent
which means that the three-body component of the force acting on atom (1) points towards atom (2) if atom (1) is closer to atom (2), i.e., $\mathcal{F}^{(1,x)}_{123}<0$ if $x_1 < x_3/2$; accordingly, it points towards atom (3) if atom (1) is closer to atom (3), i.e., $\mathcal{F}^{(1,x)}_{123}>0$ if $x_1 > x_3/2$. This feature for the linear configuration is opposite to our results for the three-body component of the CCF acting on colloid (1) for $(+,+,+)$ BCs, which is positive (i.e., repulsive) if $L_{13}<L_{12}$ and negative (i.e., attractive) if $L_{13}>L_{12}$ (not shown by a figure).

% We also compare our results with those obtained for a system consisting of three aligned intruders immersed in a granular fluid (see Ref.~\cite{PhysRevLett.108.198001} and Subsec.~\ref{subsection_L}). The authors found that the three-body component of the total force acting on the intruder analogous to \textit{colloid (3)} in our set up [see Figs.~\ref{system_sketch} and~\ref{arrangements}~(c)], contributes roughly with $25\%$ to the total force acting on that intruder. Moreover, the three-body component for a configuration corresponding to $\Lambda_{13}=1$ and $\Lambda_{12}=2$ is attractive, i.e., it pulls intruder (3) towards the other two intruders. In the case of CCFs we found that the relative contribution of the three-body component to the total force on \textit{colloid (3)} is less than $3\%$ for the same configuration with $(+,+,+)$ BCs.

\begin{figure} % FIG.9
\begin{center}
\includegraphics[width=0.65\textwidth]{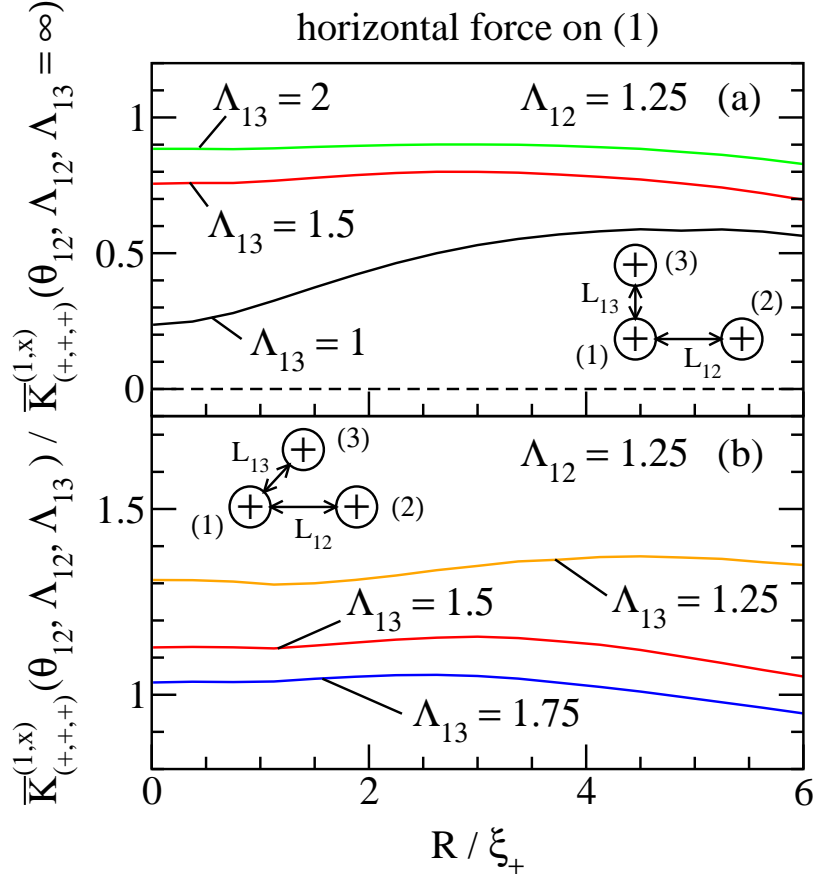}
\caption{
(Color online) Ratio between the normalized scaling function $\overline{K}^{(1,x)}_{(+,+,+)}(\Theta_{12},\Lambda_{12},\Lambda_{13})$ of the total horizontal CCF acting on colloid (1) and the normalized scaling function $\overline{K}^{(1,x)}_{(+,+,+)}(\Theta_{12},\Lambda_{12},\Lambda_{13}=\infty)$ of the pairwise CCF between the colloids (1) and (2), for $\Lambda_{12}=1.25$. We consider the configuration of a right-angled triangle in (a) and of an isosceles triangle in (b). The scaling functions are shown as functions of the scaling variable ratio $R/\xi_+=\Theta_2/\Lambda_{12}$. The black, orange, red, blue, and green curves correspond to $\Lambda_{13}=1, \Lambda_{13}=1.25,\Lambda_{13}=1.5, \Lambda_{13}=1.75,$ and $\Lambda_{13}=2$, respectively. Upon construction these ratios reduce to 1 in the limit $\Lambda_{13}\to\infty$. For not too small values of $\Lambda_{13}$ these ratios exhibit only a weak temperature dependence. The circles and arrows are \textit{schematic} representations of the colloids and of the surface-to-surface distances between them; accordingly they are not drawn to scale.}
\label{ideia_fraca}
\end{center} 
\end{figure}

In order to facilitate a comparison between our MFT-based results for cylinders in $d=4$ and possible future results for spherical colloids in $d=3$, in Fig.~\ref{ideia_fraca} we show the ratio between the normalized scaling function $\overline{K}^{(1,x)}_{(+,+,+)}(\Theta_{12},\Lambda_{12},\Lambda_{13})$ of the total horizontal CCF acting on colloid (1) and the normalized scaling function $\overline{K}^{(1,x)}_{(+,+,+)}(\Theta_{12},\Lambda_{12},\Lambda_{13}=L_{13}/R = \infty)$ of the pairwise CCF between the colloids (1) and (2) as function of $R/\xi_+ =\Theta_{12}/\Lambda_{12}$, for $\Lambda_{12}=1.25$ and several values of $\Lambda_{13}$. We consider two configurations: a right-angled triangle in Fig.~\ref{ideia_fraca} (a) and an isosceles triangle in Fig.~\ref{ideia_fraca} (b). If this mean-field expression for the ratio $\overline{K}^{(1,x)}_{(+,+,+)}(\Theta_{12},\Lambda_{12},\Lambda_{13})/\overline{K}^{(1,x)}_{(+,+,+)}(\Theta_{12},\Lambda_{12},\Lambda_{13}=\infty)$ is multiplied by a bona fide expression for the pairwise CCF in $d=3~-$  inferred from experiments or simulations or via the Derjaguin approximation from the parallel plate geometry $-$ one obtains an approximate prediction for the scaling function $\overline{K}^{(1,x)}_{(+,+,+)}(\Theta_{12},\Lambda_{12},\Lambda_{13})$ in $d=3$. Actually, this approach can be used for any of the scaling functions presented here. Figure~\ref{ideia_fraca} reveals that these ratios exhibit only a weak temperature dependence which makes it easier to implement this approach. In the case of experiments encompassing binary liquid mixtures of water and lutidine with a lower critical point at $T_c = 307$\,K and with $\xi_0^+ = 0.2$\,nm~\cite{nature_letters,epnews,PhysRevE.80.061143,PhysRevLett.101.208301,EPL:troendle.2009,trondle:074702,MolPhys:troendle.2011}, for spherical colloids of radius $R=100$\,nm in $d=3$ the temperature differences $\Delta T=|T-T_c|=0.05$\,K and $\Delta T=0.3$\,K correspond to $R/\xi_+ =2.05$ and $R/\xi_+ =6.35$, respectively. The experimental procedure described in Ref.~\cite{PhysRevLett.101.208301} allows one to stabilize temperatures close to $T_c$ within 0.01\,K over several hours.

%*******************************************

%*******************************************

\subsection{Influence of three-body CCFs on colloidal phase transitions}

Dang et al. have used effective pair potentials between colloids, as inferred from experiments, in order to perform Monte Carlo simulations of the phase behavior of colloidal suspensions immersed in near-critical solvents~\cite{jcp:1.4819896}. They have observed a shift of the calculated colloidal gas-liquid coexistence curve towards the mixed, disordered phase of the fluid as compared to the experimental data, where colloidal gas and liquid phases refer to phases poor and rich in colloids, respectively. These authors have hypothesized that this shift is due to the neglect of repulsive many-body forces in the Monte Carlo simulations and that significant improvement of the theoretical approach may be obtained already with the inclusion of three-body interactions. A similar result has been obtained by calculating the phase diagram within the framework of the random phase approximation for an effective one-component system~\cite{nova2}.

Edison et al. have investigated colloidal phase transitions in a near-critical solvent by computer simulations of an explicit model of a ternary mixture 
consisting of hard discs and two types of solvent molecules on a two-dimensional 
square lattice~\cite{nova3}. They have observed that simulations of colloids interacting 
solely via two-body forces lead to a shift of the colloidal gas-liquid coexistence 
curve towards the mixed phase as compared to the simulation results for the 
actual ternary mixture.

Our theoretical analysis provides an explanation of the observed overestimation of the width of the colloidal gas-liquid coexistence region as obtained within the framework of effective pair-potential descriptions in terms of predominantly repulsive three-body CCFs. We have shown that for geometrical arrangements of equilateral or right-angled triangles the sign of the three-body CCF is opposite to the one of the attractive two-body CCF for (+, +, +) BCs which correspond to the BCs of the aforementioned experimental and theoretical studies~\cite{jcp:1.4819896,nova2,nova3}. Moreover, the contribution of the three-body CCF to the total force is rather large if the binary liquid mixture is close to its critical point as it is apparent from Fig. 7. Of course it is technically not feasible to compute 
numerically the three-body CCF for all possible geometrical arrangements of three 
colloids, but on the basis of the presented and additional numerical results we 
conclude that the three-body CCF for (+, +, +) BCs is predominantly repulsive. 
Taking into account this repulsive force in studies of colloidal phase transitions 
in a near-critical solvents is likely to lead to better agreement with experimental 
data as compared to descriptions in terms of effective pair potentials.

In view of future experimental and theoretical studies of the phase behavior of 
non-spherical colloids in near-critical solvents we emphasize that the isotropic 
colloidal liquid phase is stabilized by the repulsive three-body CCF, while 
nematic colloidal liquids and smectic phases are disfavored, similar to results 
of earlier studies concerning the influence of interaction potentials on phase diagrams 
of fluids consisting of non-spherical particles~\cite{nova4}. On the basis of our earlier 
theoretical studies on the alignment of elongated non-spherical colloids near 
homogeneous~\cite{kondrat:204902} or chemically patterned substrates~\cite{C3SM52858H} we conclude that the 
attractive two-body CCF favors colloidal nematic liquid and smectic phases.

%-------------------------------------------------------------------------------

\section{Conclusions and discussion}\label{section_conclusion}

%-------------------------------------------------------------------------------

We have studied critical Casimir forces (CCFs) for a system composed of three equally sized, parallel cylindrical colloids of radius $R$ immersed in a near-critical binary liquid mixture (see Figs.~\ref{system_sketch} and~\ref{arrangements}). By denoting the set of boundary conditions (BC) of the system as $(a_1,a_2,a_3)$, where $a_i$ corresponds to the BC at the surface of colloid $i=1,2,3$, we have focused on the horizontal force acting on one of the colloids [labeled as ``colloid (1)''; see Figs.~\ref{system_sketch} and~\ref{arrangements}] for several geometrical configurations of the system and various combinations of BCs at the surfaces of the colloids (2) and (3), while keeping $(a_1=+)$ fixed. We have considered three distinct configurations for the three colloids (see Fig.~\ref{arrangements}): isosceles triangles, right-angled triangles, and lines.

% an isosceles triangle (i.e., $L_{13}=L_{23}$), an $L$-shape with $(L_{23}+2R)^2=(L_{12}+2R)^2 + (L_{13}+2R)^2$, and a line with $L_{23}=L_{12} + L_{13} + 2R$.

The horizontal CCF is characterized by a universal scaling function [see Eq.~\eqref{def_scaling_1}], which has been studied in the one-phase region of the solvent as function of $R/\xi_+$, where $\xi_+$ is the bulk correlation length of the binary mixture in the mixed phase. We have used mean-field theory together with a finite element method in order to calculate the order parameter profiles, from which the stress tensor yields the normalized scaling functions associated with the CCFs.

First, we have addressed the case in which the colloids are arranged in such a way that their centers correspond to the vertices of an isosceles triangle [see Fig.~\ref{arrangements}~(a)]. For the scaling function of the horizontal CCF acting on colloid (1) with $(a_1=+)$ BC, in the presence of colloids (2) and (3) with $(a_2 = +)$ and $(a_3 = -)$, respectively, we have found (Fig.~\ref{triang__ppm_L12-fixed}) that the scaling function changes sign from negative to positive, i.e., from attraction to repulsion, for fixed values of $R/\xi_+$ and of $\Lambda_{12}=L_{12}/R$ as the distance $L_{13}$ between the colloids (1) and (3) increases, signaling the occurrence of a mechanically \textit{unstable} equilibrium configuration characterized by a vanishing horizontal force. For the same set of BCs, i.e., $(+,+,-)$, we have found that for fixed values of $R/\xi_+$ and of $\Lambda_{13}=L_{13}/R$, the scaling function also changes sign upon increasing the distance $L_{12}$ between the colloids (1) and (2) (Fig.~\ref{triang__ppm_L13-fixed}). We have also found that for certain fixed geometrical parameters it is possible to observe a change of sign in the scaling function upon varying the temperature [see, for example, the black curve in Fig.~\ref{triang__ppm_L12-fixed}~(a) and the green one in Fig.~\ref{triang__ppm_L13-fixed}~(b)]. Similar results have been obtained for the scaling function in the case of $(+,-,+)$ BCs with $L_{13}$ fixed and several values of $L_{12}$ (Fig.~\ref{triang__pmp_L13-fixed}). However, for this combination of BCs the equilibrium configuration of colloid (1) is \textit{stable} in the horizontal direction.

By calculating the pairwise colloid-colloid forces and subtracting them from the total force, we have extracted the three-body component of the force acting on colloid (1). For the scaling function associated with the horizontal three-body CCFs for the configuration of isosceles triangles with $(+,+,-)$ BCs we have found a change of sign at fixed temperature (i.e., for fixed values of $R/\xi_+$) upon varying the distance between the colloids (1) and (3), while keeping the distance between the colloids (1) and (2) fixed (Fig.~\ref{triang__ppm_TB_L12-fixed}). This indicates that, for a given temperature, there is a geometrical configuration for which the three-body CCF acting on colloid (1) is zero, in which case the sum of pairwise forces provides a quantitatively reliable description of the interactions of the system. As expected, we have found that the contribution of the three-body CCF to the total force is large if the colloid-colloid distances are small, as well as if the binary liquid mixture is close to its critical point.

We have compared our results with corresponding ones for the van der Waals interaction between three atoms as described by the Axilrod-Teller three-atom potential given by  Eq.~\eqref{axilrod_teller_3bpot}. In the case of an equilateral triangle the horizontal three-atom force acting on atom (1) is positive, which implies repulsion from atom (2). This result agrees qualitatively with our results for the horizontal three-body component of the CCF acting on colloid (1), which is also repulsive for all BCs being equal. We have also compared our results for this geometry with experimental results for three-body interactions among equally sized charged colloids obtained by Brunner et al.~\cite{PhysRevLett.92.078301} for a noncritical solvent. These authors found that the three-body component of the force acting on the colloid corresponding to colloid (1) in Fig.~\ref{arrangements}~(a) is always attractive, which contrasts with our results for the horizontal three-body component of the CCF acting on colloid (1), which can be either attractive or repulsive depending on the geometrical configuration of the colloidal particles. 

Next, we have considered the configuration of right-angled triangles formed by the colloids [see Fig.~\ref{arrangements}~(b)]. The scaling function for $(+,+,+)$ BCs indicates that, keeping the geometry fixed, there is a change of sign in the lateral CCF acting on colloid (1) upon changing temperature [Fig.~\ref{L__ppp_L12-fixed}~(a)]. As expected, we have found that the three-body component of the total horizontal CCF decreases upon increasing the distance between the colloids (1) and (3) [Fig.~\ref{L__ppp_L12-fixed}~(b)]. We have compared our results with corresponding ones for the van der Waals interaction between three atoms forming the same configuration. The horizontal three-body component of the force acting on atom (1) is positive, which corresponds to repulsion from atom (2) in horizontal direction. We have observed this feature also for the three-body CCF [Fig.~\ref{L__ppp_L12-fixed}~(b)]. Moreover, we have compared our results with those obtained by molecular dynamics simulations of driven granular mixtures~\cite{PhysRevLett.108.198001}, for which it has been found that the three-body component of the force acting on an intruder, corresponding to colloid (1), is purely attractive and its relative contribution $\delta$ [see Eq.~\eqref{relative_contribution}] is $3\%$ for the case corresponding to $\Lambda_{12}=\Lambda_{13}=1$ [see Fig.~\ref{arrangements}~(b)]. This is in contrast with our results for the three-body component of the CCF which is repulsive with $\delta=24\%$ in the case $\Lambda_{12}=1.25$ and $\Lambda_{13}=1$ (black curve in Fig.~\ref{L__ppp_L12-fixed}).

In addition, we have considered the configuration in which the three colloids are horizontally aligned [Fig.~\ref{arrangements}~(c)]. In the case of $(+,+,+)$ BCs, if $\Lambda_{12}=\Lambda_{13}$ (i.e., with the center of colloid (1) being the midpoint of the line connecting the centers of colloids (2) and (3); see the red curve in Fig.~\ref{line__ppp_L12-fixed}) the CCF acting on colloid (1) vanishes due to symmetry. We have compared our results with corresponding ones for the van der Waals interaction between three aligned atoms in which case the three-body component of the force acting on atom (1) can be either attractive or repulsive, depending on whether atom (1) is closer to atom (2) or atom (3), respectively. This is opposite to our results for the three-body component of the CCF acting on colloid (1). 

Finally, in order to suggest a method to accomplish a future quantitative comparison between our MFT-based results for cylinders in $d=4$ and possible results for spherical colloids in $d=3$, in Fig.~\ref{ideia_fraca} we have shown the ratio between the normalized scaling function $\overline{K}^{(1,x)}_{(+,+,+)}(\Theta_{12},\Lambda_{12},\Lambda_{13})$ of the total horizontal CCF acting on colloid (1) and the normalized scaling function $\overline{K}^{(1,x)}_{(+,+,+)}(\Theta_{12},\Lambda_{12},\Lambda_{13}=L_{13}/R=\infty)$ of the pairwise CCF between the colloids (1) and (2). As function of $R/\xi_+ = \Theta_2/\Lambda_{12}$ and for two configurations this ratio exhibits only a weak temperature dependence. Multiplying this ratio as obtained within MFT with the proper $3d$ counterpart of the denominator $-$ taken from experiment, simulation, or from the slab geometry via the Dejarguin approximation $-$ renders an approximate prediction for the numerator in $d=3$. This approximate expression for the numerator yields its correct expression in $d=3$ for the limiting value of the scaling variable as taken for the denominator within the ratio. This scheme can be applied to any of the scaling functions considered here.

% \vspace{2cm}
% \textcolor{red}{\texttt{The next paragraph will be rewritten as soon as numerical results for the force on colloid (3) are available.\\ \\}}
% 
% \textcolor{red}{\textit{
% We also compare with results obtained for driven granular mixtures~\cite{PhysRevLett.108.198001}. The authors calculate the force acting on the intruder which is equivalent to colloid (3) in our system [see Fig.~\ref{arrangements}~(c)] and show that the relative contribution from the three-body component is attractive and corresponds to $25\%$ of the total force, while our results for the CCF, as mentioned before, is less than $3\%$ and is less than our numerical accuracy.}}

%-------------------------------------------------------------------------------

% \bibliography{biblio}

\end{document}